# Distributed Market Clearing Approach for Local Energy Trading in Transactive Market

Mohsen Khorasany, Yateendra Mishra, Gerard Ledwich
School of Electrical Engineering and Computer Science,
Queensland University of Technology,
Brisbane, Australia
m.khorasany@qut.edu.au, yateendra.mishra@qut.edu.au, g.ledwich@qut.edu.au

*Abstract*— **This paper proposes a market clearing mechanism for energy trading in a local transactive market, where each player can participate in the market as seller or buyer and tries to maximize its welfare individually. Market players send their demand and supply to a local data center, where clearing price is determined to balance demand and supply. The topology of the grid and associated network constraints are considered to compute a price signal in the data center to keep the system secure by applying this signal to the corresponding players. The proposed approach needs only the demanded/supplied power by each player to reach global optimum which means that utility and cost function parameters would remain private. Also, this approach uses distributed method by applying local market clearing price as coordination information and direct load flow (DLF) for power flow calculation saving computation resources and making it suitable for online and automatic operation for a market with a large number of players. The proposed method is tested on a market with 50 players and simulation results show that the convergence is guaranteed and the proposed distributed method can reach the same result as conventional centralized approach.**

*Index Terms*-- **Transactive Energy, Electricity Market, Distributed optimization, Energy Exchange.**

## I. Introduction

With increasing penetration of distributed energy resources (DERs), a huge number of DER owners have the capability to participate in the residential electricity market by managing their own generators and loads. The integration of distributed resources in the market clearing requires a platform which enables them to actively participate in a transactive market. The transactive market is the concept of linking supply and demand in the electricity system primarily through the response of electric loads and generation on the consumer side of the meter to price signals [1]. In this market, interactive controllers are used which can react to the price changes, returning information back to the central controller, and automatically acting on behalf of the end-user load [2].The study of market design for distribution systems has received high research momentum in the recent years and different approaches have been used for clearing the market.

Generally, market clearing in the transactive market can be performed using centralized or distributed approaches. In a centralized market clearing, local decisions are usually made centrally by having information of all market players in the central controller [3]. The centralized algorithms contribute to heavy computational burden and are not appropriate for a large number of market players [4]. On the other side, distributed approaches can be used to find the global optimal decision by allowing local agents to iteratively share information through two-way communication links [5]. In these approaches, all players reach an agreement when they agree upon the value of the shared information [6]. Different quantities and control signals can be applied as shared information among market players such as voltage, estimated power mismatches, and market price. By applying distributed method, the computation and control can be distributed across the grid.

Recently, there is an increasing interest in the literature on distributed algorithms [7-11]. A distributed method for negawatt trading is discussed in [7], where a market algorithm is presented to solve energy trading in the real-time electricity market. This paper only focuses on adjustment of power imbalance caused by generating error without considering energy trading among market players. Authors in [8] deal with distributed demand response program in the multiseller-multibuyer environment, where the proposed approach does not consider topology of the grid and network constraints in energy trading. A game-theoretic framework for a next-generation retail electricity market with high penetration of distributed residential electricity suppliers is presented in [9]. The proposed approach needs private information of all market players to solve the social welfare maximization problem in the market clearing. Authors in [11] address the interaction among multiple sellers and buyers in the grid by using a data center as a coordinator among players. In this paper, the data center calculates the clearing price without considering network constraints.

In the most of the works on distributed market clearing in the distribution network, the topology of the grid or privacy of the market players' information is neglected. In this paper, a distributed market clearing for energy trading in the transactive market is proposed inspired by the presented algorithms in [10] and [11]. In this approach, market players send their demand and supply to a local data center, where market clearing price is calculated to balance demand and supply. Also, the topology of the grid is considered in the market clearing by computing a price signal in the data center and applying this signal to the corresponding players to keep the system secure. The proposed approach needs only the demanded/supplied power to reach global optimum which means that utility and cost function parameters would remain private for each player. Also, the computational burden in this approach would be lower compared to central computation which makes it suitable for a market with a large number of players.

This paper contributes to propose a local market platform for energy trading in the transactive environment using local area communication network as communication architecture. Also, the design of a distributed market clearing approach for energy trading considering topology of the grid and network constraints is presented. The proposed approach uses distributed computation and direct load flow approach to reduce computation resources which make it suitable for online operation in a market with numerous number of market players.

## II. SYSTEM MODEL

### A. Assumptions

In this market, sellers and buyers participate in a local market to trade energy for the next time slot through a data center and distributed agents which act on behalf of market players and send information and automatically adjust settings in response to a coordination signal. It is assumed that market players are automated trading agents that can predict their future power production and consumption based on historical data and do this prior trading time slot to participate in the market. Market players send their demand and supply to the data center, where market clearing price is calculated as a centralized signal in order to balance demand and supply in the market. Therefore, an appropriate communication and messaging architecture is required among market players and data center.

The communication architecture can be implemented using wireless communication in the existing grid, where market players are equipped with a device such as Raspberry Pi and connected to each other and data center through local area communication network. The Raspberry Pi is a credit-card-sized computer that can be used for various purposes such as metering and communicating. In recent years, there are several projects which work on introducing a new module to extend Raspberry Pi as a smart meter with the capability of communication. Authors in [12] proposed a cheap and easy way to use Raspberry Pi for automatic metering system which can obtain the usage data and transmit the acquired data to other devices. SmartPi in one of the other modules which turn Raspberry Pi into a smart meter [13]. The Smart Pi extends the Raspberry Pi to measure voltage and current in a contactless manner. All measurement data can be stored and are accessible via the local network or internet for disposal. Using SmartPi with wireless communication provides two-way communication for market players and can be implemented with low investment cost. Also, by dividing the whole grid to several local markets, the distance among market players would be low and wireless communication can provide required bandwidth.

### B. Problem formulation

The total grid is divided into several areas and in each area, there is a local market with a set of sellers indexed by $i$, where $i \in \mathbb{N}_S \triangleq \{1,2,...,N_S\}$ and a set of buyers indexed by $j$, where $j \in \mathbb{N}_B \triangleq \{1,2,...,N_B\}$. In each time slot, market players join to the market by sending their required/surplus energy to the data center. Each player has its own objective in the market and tries to maximize its welfare. The $i^{th}$ seller objective ($SO_i$) is to maximize social welfare and can be modelled using (1).

$$SO_i: \max_{s_i^{\min} \leq s_i \leq s_i^{\max}} s_i(\wp + \varpi) - C_i(s_i) \quad (1)$$

where $s_i$ is the sold power in the market; $\wp$ is market clearing price; $\varpi$ is the voltage management price signal and $C_i$ is cost function of seller which is modelled by a quadratic cost function as (2).

$$C_i(s_i) = a_i s_i^2 + b_i s_i + \gamma_i \quad (2)$$

where $a_i$, $b_i$, and $\gamma_i$ are predetermined parameters and are assumed to be private for each player. The $j^{th}$ buyer objective ($BO_j$) is the social welfare maximization and can be modelled in the same way using (3), where utility of buyer for consuming the demanded power is modelled using (4) [10].

$$BO_j: \max_{d_j^{\min} \leq d_j \leq d_j^{\max}} U_j(d_j) - d_j(\wp + \rho) \quad (3)$$

$$U_j(d_j) = \begin{cases} \omega_j d_j - \delta_j d_j^2, & d_j < \omega_j/2\delta_j \\ \omega_j^2/2\delta_j, & d_j \geq \omega_j/2\delta_j \end{cases} \quad (4)$$

where $d_j$ is the bought power by buyer $j$ in the market; $\rho$ is the line flow management price signal; $U_j$ shows utility of buyer for consuming this power; and $\omega_j$ and $\delta_j$ are predetermined parameters that are considered to be private for each buyer. In this market, each individual player tries to maximize its welfare selfishly. However, from social fairness point of view, the objective is to maximize the total welfare of all market players. Thus, the total objective ($TO$) in the market can be written as (5)

$$TO: \max_{S,D} \left( \sum_{j=1}^{N_B} U_j(d_j) - \sum_{i=1}^{N_S} C_i(s_i) \right) \quad (5)$$

where $\mathbf{D} \triangleq \{d_j | j \in N_B\}$ is demand vector of buyers, and $\mathbf{S} \triangleq \{s_i | i \in N_S\}$ is the supply vector of sellers. The data center aims to clear the market to match supply and demand. Thus, the total demanded energy by buyers should be equal to the total supplied energy by sellers at the end of market clearing as (6).

$$\sum_{i=1}^{N_S} s_i = \sum_{j=1}^{N_B} d_j \quad (6)$$

## III. MARKET CLEARING APPROACH

### A. Distributed algorithms

The optimization problem in (5) is a convex problem subject to affine constraints where the inequality constraints related to minimum and maximum limits of demand and supply are local and can be treated as the boundaries of the domain of

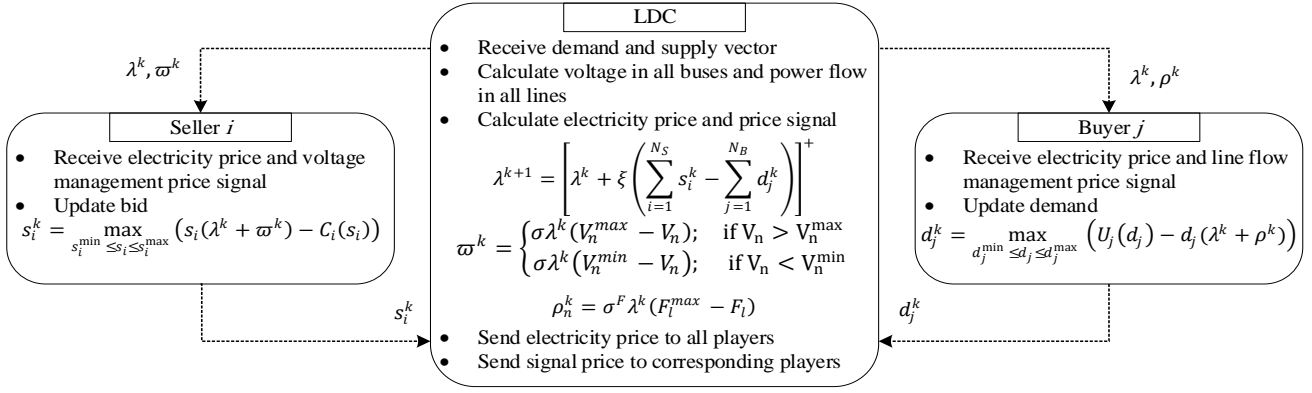
Fig. 1. Information flow and computation procedure

the problem. The problem in (5) can be augmented as (7) using Karush-Kuhn-Tucker (KKT) multipliers.

$$R = \sum_{j=1}^{N_B} U_j(d_j) - \sum_{i=1}^{N_S} C_i(s_i) + \lambda\left(\sum_{i=1}^{N_S} s_i - \sum_{j=1}^{N_B} d_j\right) \quad (7)$$

where $\lambda$ represents Lagrangian or KKT multiplier and is the same as market clearing price ($\wp$). By applying dual decomposition, a distributed iterative approach can be developed to maximize $R$ without any need to have individual parameters of all market players. The distributed updating rules for demand, supply and price, derived by utilizing the primal-dual gradient descent method [14] are presented in (8), (9), and (10) respectively.

$$d_j^k = \max_{d_j^{min} \le d_j \le d_j^{max}} \left(U_j(d_j) - d_j(\lambda^k + \rho^k)\right) \quad (8)$$

$$s_i^k = \max_{s_i^{min} \le s_i \le s_i^{max}} \left(s_i(\lambda^k + \varpi^k) - C_i(s_i)\right) \quad (9)$$

$$\lambda^{k+1} = \left[\lambda^k + \xi\left(\sum_{i=1}^{N_S} s_i^k - \sum_{j=1}^{N_B} d_j^k\right)\right]^+ \quad (10)$$

where $k$ is index of iteration, $\xi$ denotes step size and the notation $[.]^+$ denotes max $\{.,0\}$.

*B. Price signal for network constraints*

In this paper, the data center is responsible to ensure the network constraints are considered in the market clearing process. Once **D** and **S** are received, voltage and power flow in the system should be calculated using direct approach [15], where a distribution load flow can be obtained by solving (11-13).

$$I_n = \left(\frac{P_n + jQ_n}{V_n}\right)^* \quad (11)$$

$$\lfloor \Delta V \rfloor = \lfloor DLF \rfloor \lfloor I \rfloor \quad (12)$$

$$\lfloor V \rfloor = \lfloor V_0 \rfloor + \lfloor \Delta V \rfloor \quad (13)$$

where $n$ is index of nodes in the network, $I_n, V_n, P_n$ and $Q_n$ are injection current, voltage, active and reactive power in node $n$ respectively. Voltage and current in all nodes are presented by matrix $\lfloor V \rfloor$ and $\lfloor I \rfloor$ respectively. The direct load flow (DLF) matrix is obtained by (14) using the branch current to node voltage matrix (BCBV) and the node-injection to branch-current matrix (BIBC). These two matrices are developed based on the structure of distribution system using line impedances [15]. The $\lfloor B \rfloor$ represents the line current matrix as in (15).

$$\lfloor DLF \rfloor = \lfloor BIBC \rfloor \lfloor BCBV \rfloor \quad (14)$$

$$\lfloor B \rfloor = \lfloor BIBC \rfloor \lfloor I \rfloor \quad (15)$$

The power flow in each line ($F_l$) is obtained using the line current and node voltage. The presented market is designed for the active power trading and $Q_n$ is assumed to be zero in (11). The DLF matrix is a unique and constant matrix for each topology. Therefore, the data center can easily calculate power flow and voltage in different nodes by having injected and withdrawn power in each node. The active power ($P_n$) in each node ($n$) is obtained by (16).

$$P_n = \begin{cases} d_j & \text{if player is a buyer} \\ -s_i & \text{if player is a seller} \end{cases} \quad (16)$$

The obtained voltage for each node and flow in each line are bounded by (17) and (18) respectively.

$$V_n^{min} \le V_n \le V_n^{max} \quad (17)$$

$$F_l^{min} \le F_l \le F_l^{max} \quad (18)$$

where $V_n^{min}, F_l^{min}$ and $V_n^{max}, F_l^{max}$ are minimum and maximum voltage and line flow limits at node $n$ and line $l$ respectively. A price signal is considered to model system topology and keep the voltage and line flow of the system in the desirable range. At each iteration, if there is a voltage increase/decrease at a node, a voltage management price signal is sent by data center to the nearest seller to decrease/increase its generation. The reason for considering the nearest generation for voltage balance is to reduce power losses in the lines. This price acts as a control signal for sellers and is calculated using (19).

$$\varpi_n^k = \begin{cases} \sigma^v \lambda^k (V_n^{max} - V_n); & \text{if } V_n > V_n^{max} \\ \sigma^v \lambda^k (V_n^{min} - V_n); & \text{if } V_n < V_n^{min} \end{cases} \quad (19)$$

where $\sigma^v$ is the price adjustment coefficient for voltage. Congestion in the line is alleviated by sending a price signal to the buyers using that line to reduce their demand. Power Transfer Distribution Factor (PTDF) is utilized to indicate the buyers who use that particular line [16], [17]. PTDF is an approximation of the first order sensitivity of the active power flow and represents change in active power flow over certain line, caused by change in active power generation in certain node. At each iteration, the line flow constraint is checked and if there is an overload or congestion in the line, the price signal is sent to the buyers who use that line to reduce their demand. The power flow price signal is calculated by (20).

$$\rho_n^k = \sigma^F \lambda^k (F_l^{max} - F_l) \quad (20)$$

where $\sigma^F$ is the price adjustment coefficient for line flow. In this market, all monetary transfers must be performed between buyers and sellers i.e. $\sum_n \varpi_n = \sum_n \rho_n$. Therefore, in each iteration $\sigma^v$ and $\sigma^F$ should be adjusted in a way that resolve this constraint (Fig. 1). Once **D** and **S** are received, the data center checks voltage constraint and line congestion and calculates market price ($\lambda$) and price signals ($\varpi, \rho$) for market clearing and sends them back to the market players and waits for updated demand and supply. Market players update their demand and supply using (8) and (9). This algorithms repeat until demand and supply are converged.

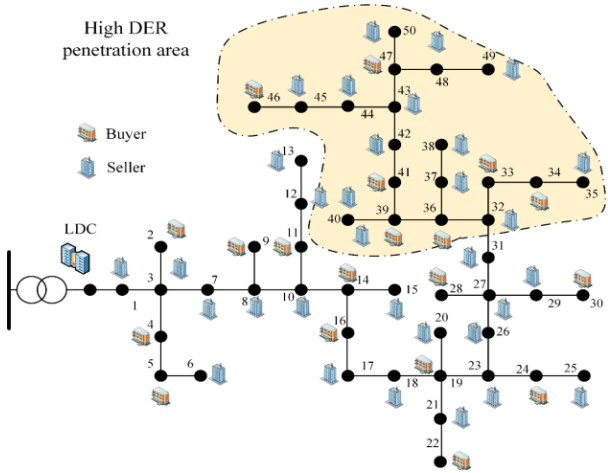

Fig. 2. Schematic of the test system with 50 players

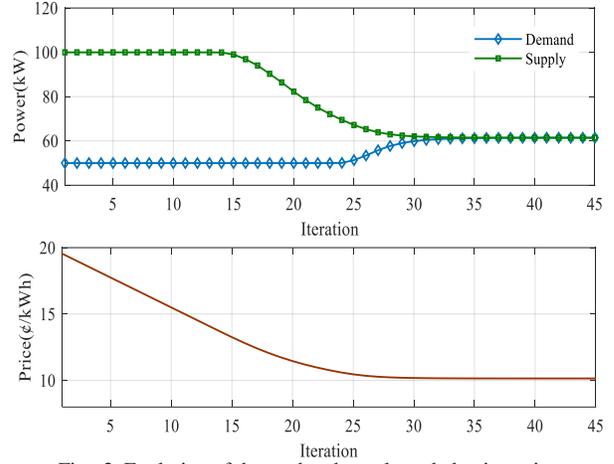

Fig. 3. Evolution of demand and supply and clearing price

## IV. CASE STUDIES

The performance of the proposed approach for market clearing with 50 players (25 sellers and 25 buyers as shown in Fig. 2) is evaluated. The utility function parameters of buyers ($\omega_j$ and $\delta_j$) are selected randomly from the interval [0, 0.9] and [13, 17] respectively, whereas the cost function parameters for sellers ($a_i$ and $b_i$) are selected randomly from the interval [0, 0.9] and [3, 8] respectively and $\gamma_i$ is set to 0. The minimum and maximum of demand and supply of buyers and sellers are [2, 4] kW. For ease of illustration, the market clearing for one time slot is considered, where all of the market players send their demand and supply to the data center to trade energy in the local market.

The convergence of demand-supply and evolution of electricity price are shown in Fig. 3 and shows that the power balance condition is satisfied. The coordination parameter for market players is $\lambda^0 = 20$ (¢/kWh) and all sellers try to sell the maximum power with this price to earn more welfare. However, since this price is not optimal for buyers and reduces their welfare, they try to buy the minimum power with this price. Therefore, there is a mismatch between demand and supply at this price. The data center updates the price based on this mismatch and sends it back to the players. Once price decreases, buyers increase their demand and sellers decrease their supply till demand and supply are converged at price $\lambda = 10.12$ (¢/kWh). The corresponding execution time of the algorithm is about 1.95 seconds.

In this paper, the topology of the grid and associated network constraints are used to compute a price signal in the data center to keep the system secure by applying this signal to the corresponding players. Fig. 4 presents voltage in different nodes with and without considering price signal to illustrate impact of considering network topology. The results demonstrate that without considering the price signals and network topology in the market clearing process, the voltage in nodes numbers 36 to 50 is higher than 1 p.u. These voltages are reduced when price signal is applied to the corresponding players and are in the desired range. Also, since the power from each seller is allocated to the nearest buyers, the power flow in all lines are lower than the maximum line capacity. Thus, the proposed approach can preserve the network stability by considering network constraints in the market clearing.

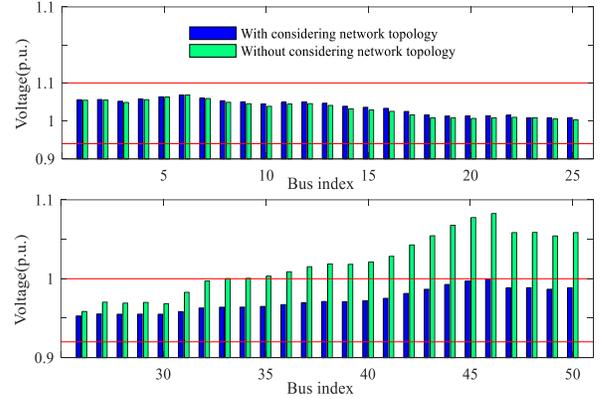

Fig. 4. Voltage profiles of different buses with and without considering network topology

The proposed approach for market clearing is implemented for several markets with different number of players. The market clearing price for different numbers of sellers in the market is extracted to investigate the impact of DERs integration in the grid on the electricity price. In this case, it is assumed that the total number of players in the market is fixed (50 players) and the number of players who have DER and participate in the market as seller changes. Results are shown in Fig. 5(a-b), which illustrate that by increasing the number of sellers and providing more generation in the market the final market clearing price is decreased. Also, as the number of sellers is increased, the convergence rate of demand and supply increases and number of iterations decreases, which means that in a market with fixed number of players, market clearing process for a higher number of sellers needs less time.

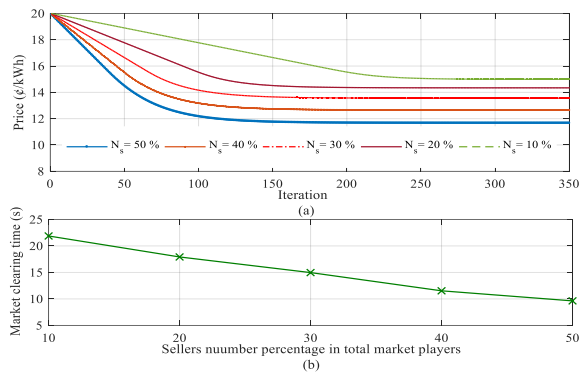

Fig. 5. Market clearing price (a) and running time (b) for markets with different number of sellers

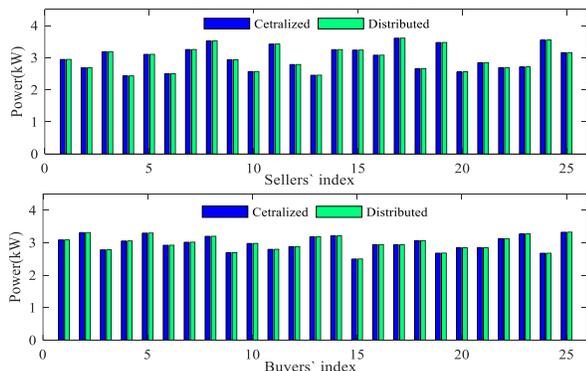

Fig. 6. Demand and supply vector in centralized and distributed approach

The optimization problem in (5) can be solved in a central computation center and by having individual information of all market players. However, the proposed method in this paper can solve this problem by distributed computation and without any need to revealing individual information of players. The demand and supply vectors (**D** and **S**) which are the solution of the optimization problem, are compared in centralized and distributed method to validate the results of distributed method. The results are shown in Fig. 6 and verify that the proposed method can reach the same results as when using a centralized computation.

## V. CONCLUSIONS

In this paper, a distributed market clearing approach for local energy trading in the transactive market is presented. The proposed approach clears the market without any need for private information of market players and considers the topology of the grid and network constraints in the market clearing. Utilizing a local data center and distributed agent methodology the computational burden in this approach would be lower which makes it suitable for a market with large number of players. Market players send their demand and supply to the data center and market clearing price is calculated as a coordination signal to balance demand and supply. Also, a price signal is considered to keep voltages in nodes and flow in lines in the limit. Simulation results verified that the proposed approach can balance demand and supply in a distributed manner considering network constraints and reaches the same results as centralized market.

For the future work, the proposed approach will be extended to several local markets simultaneously and local data centers communicate with neighboring areas to trade energy.

## REFERENCES

[1] C. King, "Chapter 9 - Transactive Energy: Linking Supply and Demand Through Price Signals A2 - Sioshansi, Fereidoon P," in *Distributed Generation and its Implications for the Utility Industry*, ed Boston: Academic Press, pp. 189-204, 2014.

[2] J. C. Fuller, K. P. Schneider, and D. Chassin, "Analysis of Residential Demand Response and double-auction markets," *Proc. IEE PES Gen. Meet*., 2011.

[3] M. Khorasany, Y. Mishra, and G. Ledwich, "Peer-To-Peer Market Clearing framework for DERs Using Knapsack Approximation Algorithm," *Proc. IEEE ISGT Europe*, Torino, 2017.

[4] R. Mudumbai, S. Dasgupta, and B. B. Cho, "Distributed Control for Optimal Economic Dispatch of a Network of Heterogeneous Power Generators," *IEEE Trans. Power Syst.,* vol. 27, pp. 1750-1760, 2012.

[5] H. Pourbabak, J. Luo, T. Chen, W. Su, "A Novel Consensus-based Distributed Algorithm for Economic Dispatch Based on Local Estimation of Power Mismatch," *IEEE Trans. Smart Grid,* 2017.

[6] G. Binetti, A. Davoudi, F. L. Lewis, D. Naso, and B. Turchiano, "Distributed Consensus-Based Economic Dispatch With Transmission Losses," *IEEE Trans. Power Syst.,* vol. 29, pp. 1711-1720, 2014.

[7] Y. Okawa and T. Namerikawa, "Distributed Optimal Power Management via Negawatt Trading in Real-time Electricity Market," *IEEE Trans. Smart Grid,* vol. 8, Is. 6, pp. 3009-3019, Nov. 2017.

[8] R. Deng, Z. Yang, F. Hou, M. Y. Chow, and J. Chen, "Distributed Real-Time Demand Response in Multiseller-Multibuyer Smart Distribution Grid," *IEEE Trans. Power Syst.,* vol. 30, pp. 2364-2374, 2015.

[9] N. Zhang, Y. Yan, and W. Su, "A game-theoretic economic operation of residential distribution system with high participation of distributed electricity prosumers," *Applied Energy,* vol. 154, pp. 471-479, 2015.

[10] P. Samadi, H. Mohsenian-Rad, R. Schober, and V. Wong, "Advanced Demand Side Management for the Future Smart Grid Using Mechanism Design," *IEEE Trans. Smart Grid,* vol. 3, pp. 1170-1180, 2012.

[11] F. Kamyab, M. Amini, S. Sheykhha, M. Hasanpour, M. Jalali, "Demand Response Program in Smart Grid Using Supply Function Bidding Mechanism," *IEEE Trans. Smart Grid,* vol. 7, pp. 1277-1284, 2016.

[12] P. A. Chandra, G. M. Vamsi, Y. S. Manoj, and G. I. Mary, "Automated energy meter using WiFi enabled raspberry Pi," *Proc. IEEE Int. Conf. on Recent Trends in Elect., Inf. & Commun. Tech. (RTEICT)*, Bangalore, 2016.

[13] SmartPi, "Turn your Raspberry Pi into a smart meter", Emanager, [Online], Available: http://www.emanager.eu/en/products/smartpi.

[14] S. Boyd, N. Parikh, E. Chu, B. Peleato, J. Eckstein, "Distributed Optimization and Statistical Learning via the Alternating Direction Method of Multipliers," in *Found. and Trends in Mach. Learning*. vol. 3, pp. 1-122, 2010.

[15] J.-H. Teng, "A direct approach for distribution system load flow solutions," *IEEE Trans. Power Delivery,* vol. 18, pp. 882-887, 2003.

[16] M. Khorasany, Y. Mishra, and G. Ledwich, "Auction Based Energy Trading in Transactive Energy Market with Active Participation of Prosumers and Consumers," *Proc. Australasian Uni. Power Eng. Conf. (AUPEC)*, Melbourne, 2017.

[17] L. Minghai and G. Gross, "Role of distribution factors in congestion revenue rights applications," *IEEE Trans. Power Syst.,* vol. 19, pp. 802-810, 2004.